\documentclass[conference]{IEEEtran}

\usepackage{graphicx} % Required for inserting images
\usepackage{amssymb}
\usepackage{booktabs}
\usepackage{makecell}
\usepackage{wrapfig}

\title{A Survey on the Use of Partitioning in IoT-Edge-AI Applications}
\author{Guoxing Yao, Lav Gupta}
\date{June 2024}

\begin{document}
\maketitle
\footnotetext[1]{A version of this paper is currently submitted and is under review with IEEE Communications Surveys and Tutorials.}

\begin{abstract}
Centralized clouds processing the large amount of data generated by Internet-of-Things (IoT) can lead to unacceptable latencies for the end user.  Against this backdrop, Edge Computing (EC) is an emerging paradigm that can address the shortcomings of traditional centralized Cloud Computing (CC). Its use is associated with improved performance, productivity, and security. Some of its use cases include smart grids, healthcare Augmented Reality (AR)/Virtual Reality (VR). EC uses servers strategically placed near end users, reducing latency and proving to be particularly well-suited for time-sensitive IoT applications. It is expected to play a pivotal role in 6G and Industry 5.0. Within the IoT-edge environment, artificial intelligence (AI) plays an important role in automating decision and control, including but not limited to resource allocation activities, drawing inferences from large volumes of data, and enabling powerful security mechanisms. The use cases in the IoT-Edge-cloud environment tend to be complex resulting in large AI models, big datasets, and complex computations. This has led to researchers proposing techniques that partition data, tasks,  models, or hybrid to achieve speed, efficiency, and accuracy of processing.  This survey comprehensively explores the IoT-Edge-AI environment, application cases, and the partitioning techniques used.  We categorize partitioning techniques and compare their performance.  The survey concludes by identifying open research challenges in this domain.
\end{abstract}


\section{Introduction}
\noindent
\IEEEPARstart{I}{oT} provide efficient interconnection and data exchange among diverse IoT devices at the edge. Edge and IoT form efficient communication in between and overcome the challenge of handling massive and heterogeneous data. With the help of AI, IoT-EC-AI ensures timely communication and accurate analysis for real-world applications. Partitioning is necessary to arrange data, tasks, and AI models in IoT-EC-AI to improve its scalability.

\subsection{IoT drives Cloud Computing to Edge Computing-Artificial Intelligence}

The term Internet of Things (IoT)  was first introduced by Kevin Ashton in 1999 while using Radio-frequency identification (RFID). Today it represents a transformative technology that connects various devices, enabling them to communicate and provide unprecedented services to users. IoT devices encompass a wide range of technologies, including sensors, actuators, activity trackers, smartphones, and other platforms\cite{1_kong2022edge}. According to a report by Grand View Research, the global IoT market is projected to expand at a compound annual growth rate (CAGR) of 12.7\% from 2023 to 2030.  \cite{hireotter_2}. In parallel, it is estimated that by 2025, IoT devices will generate approximately 73.1 ZB (zettabytes) of data \cite{Howarth_3} of heterogeneous data. IoT technology has yielded various successful applications, including smart homes, smart cities, smart grids, smart agriculture, AR, self-driving vehicles, smart retail, and smart healthcare. Notwithstanding their versatility, IoT devices have limited computational, networking, and storage capabilities, necessitating the uploading of captured data to remote servers for advanced processing or persistent storage. At the same time, most IoT applications are sensitive to responsive time, posing new challenges for traditional Cloud Computing architectures. Consequently, opportunities and challenges presented by IoT technology have been intriguing researchers for quite some time now.

Regardless of the time-sensitive nature of IoT applications, Cloud Computing (CC) originally supplemented the Internet with high-performance processing and storage that personal devices(PC, Mobile devices) lack. Following the dot-com bubble burst in March 2000, CC was established as a novel computing infrastructure, comprising both software and hardware, designed to process data at a limited number of centralized centers globally. For instance, Google Cloud has only 40 regions to deploy their data center\cite{122_googlecloud}.  However, it is now being realized that CC alone is not suitable for IoT applications. CC excels in computation and storage but has the drawback of communication latency when it handles massive volumes of data uploading and downloading remotely. This challenge is further exacerbated by issues such as network congestion, bandwidth constraints, and reduced Quality of Experience (QoE). For instance, applications in healthcare demand a high level of QoE and low latency to support diverse tasks \cite{shukla2021improving_8}. Conversely, a huge quantity of data generated from IoT devices would inundate the CC servers and potentially strain the bandwidth of the public Internet during its uploading and downloading \cite{jamshed2022challenges_9}.
\begin{figure}[!t]
  \centering
    \includegraphics[width=0.5\textwidth]{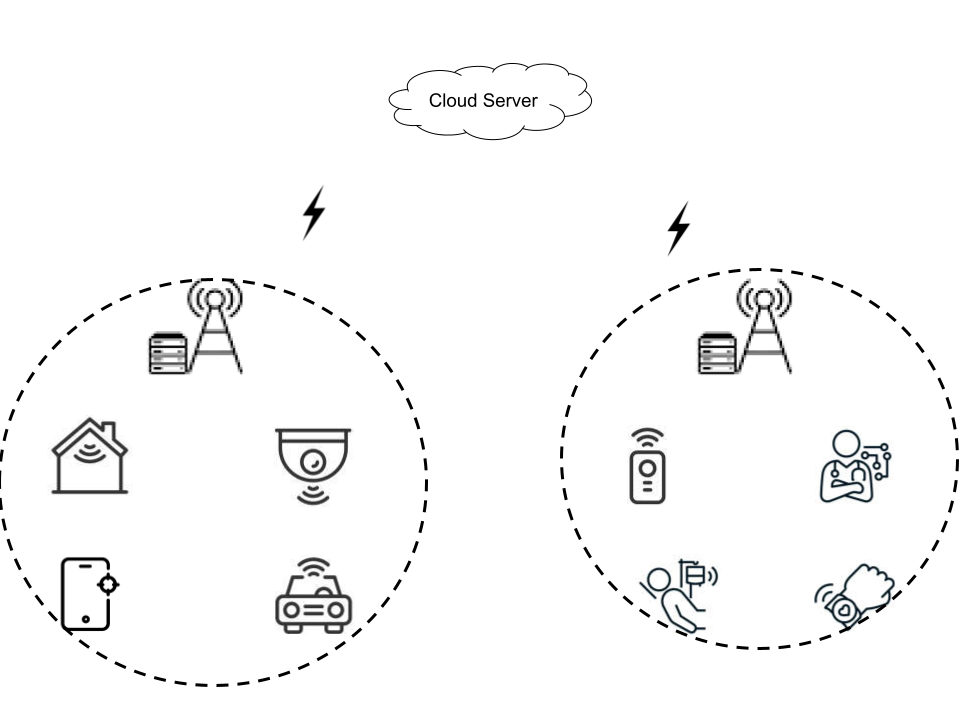}
    \caption{Architecture of IoT-EC-CC based system}
    \label{fig:1}
\end{figure}

Edge Computing (EC) is introduced as a solution to some of the challenges of CC described above. It was first conceived in the late 1990s to deliver time-sensitive content from edge servers to end-users rather than relying on remote servers. Generally, EC servers have comparatively fewer resources than CC servers, but the edge servers are strategically geographically located much closer to the data sources, which reduces transmission latency.  EC reduces the dependence on wide area networks of telecommunications service providers. The relatively shorter links between IoT devices and the edge servers ensure better network availability. Additionally, beyond addressing latency concerns, though edge servers do not own rich resources as cloud servers, a collective of edge servers can cooperate as a distributed computing paradigm for processing data to reduce process latency. Figure 1 gives an architecture of the IoT-EC systems, which circulates the IoT devices and EC servers together within a short distance, while a long distance to the cloud server.  Workloads are offloaded from IoT devices to the nearby EC servers and further to remote CC servers if they exceed the EC server's computational capacity.

In the IoT-EC environment, the heterogeneous data and the lower resource capacity of IoT and EC require better 1) resource configuration and 2) data processing techniques, leading to the increasing use of AI. AI enables automation of monitoring and management and brings predictive capabilities to the system. AI operates on heterogeneous data from IoT devices to identify underlying rules and patterns, thereby driving systems toward predefined objectives. In recent studies, it is seen from the literature that Deep Neural Network (DNN) is a promising class of AI models in IoT-EC implementations. It is usually applied through DNN models that can assist in handling data processing with high accuracy.  In addition to data processing, AI techniques have been proposed by researchers for balancing workload between IoT devices and EC servers.

\subsection{Motivation for this Survey}

The combination of IoT, EC, CC, and AI (IoT-EC-CC-AI) has emerged as a new trend, in contrast to the traditional Internet paradigm centralized on CC. This new trend proposes new challenges and drives us to answer the question: what considerations are important when designing IoT-EC-CC-AI applications? 

\subsubsection{New requirements for IoT-EC-CC-AI}
IoT-EC-AI shares a common vision for the future of the Internet, which calls for addressing the mismatch between CC and its finite number of remote data centers, and the rapid proliferation of diverse IoT devices that vie for access to EC resources. First, the challenges arising from IoT-EC-AI, particularly the vast and diverse datasets in various formats from massive IoT devices and time-sensitive applications, undermine the advantages of the traditional CC platforms due to the high data transmission latency. Secondly, this complex scenario presents intricate challenges within a physical environment constrained by the limitations of IoT devices. For instance, IoT devices typically have short-term, non-rechargeable battery life, and the premature failure of IoT batteries can have significant consequences, disrupting applications and compromising the validity of big data analysis, thereby sparking research into resource utilization and battery consumption prediction \cite{jolly2021iot_12}\cite{bhattacharya2021deep_13}. Thirdly, since IoT-EC-CC diagram has a more complex architecture, data becomes more vulnerable than the traditional CC paradigm. Another example of a challenge in IoT time-sensitive scheduling applications is providing tolerable Quality of Service (QoS) to end-users \cite{15_abolhassani2022real}. As a whole, the convergence of IoT, EC, CC, and AI has blurred the boundaries between them but has given rise to a sequence of new requirements on the Internet model\cite{singh2023edge_11}, including 1) how to store data in a distributed way, 2) how to keep data privacy, 3) how to configure task utilization, 4) how to optimize DNN model by distributing their layers in IoT-EC-CC-AI paradigm.

\subsubsection{Partitioning as the solution to complex Challenges of IoT-EC-CC-AI} 
The combination of IoT devices, EC servers, and CC servers forms a layered architecture. Meanwhile, DNNs possess a naturally multi-layered structure, making them well-suited for processing tasks of each layer of neurons in a distributed manner. In the context of this paper, DNN partitioning involves processing by two or more layers for various ends like improving training convergence, reducing latency, acquiring the ability to run complex models, or handling large heterogeneous datasets. In addition to DNN partitioning, data partitioning is employed in a large number of use cases involving the concurrent use of IoT, EC, CC, and AI. Thus, data partitioning in big data applications has been a crucial strategy employed to enhance the efficiency of data processing and analysis. As vast datasets become increasingly common, partitioning involves dividing these large datasets into smaller, more manageable subsets for parallelization of tasks and data privacy. Last, task partitioning has attracted research focus to promote the efficiency of resource allocation among IoT-EC-CC. 

In summary of these two subsections, IoT applications introduce new requirements and challenges that will potentially transform the traditional Internet paradigm. The transition from IoT-CC to IoT-EC-CC-AI is driven by the requirements for time-sensitive data processing, reduced latency, better security, and enhanced scalability. We believe partitioning is a valid and powerful solution to address the complex demands of IoT within the EC-CC and AI ecosystems. By breaking down complex applications into smaller, manageable sub-tasks, partitioning tasks and models can optimize resource utilization, and partitioning data can improve system performance, making it a key element in IoT-EC-CC-AI convergence.

\subsection{Existing Surveys and Reviews}

Several surveys and reviews have been conducted to provide valuable insight into EC, CC, IoT, and AI individually or in combination. We have selected many important ones and summarized them in Table I. Most of these cover aspects of IoT, CC, and EC. Some of these also discuss the open future research and challenges, while most of them cover some aspects of the IoT-EC-CC-AI environment. Notably, the topic of partitioning has not been extensively discussed. Below is a summary of the surveys and reviews published in the last four years. We have categorized them into EC-oriented surveys, IoT-EC-oriented surveys, and AI-oriented surveys. 

\subsubsection{EC-oriented Surveys} 

In Table I, EC has been a very popular topic among researchers during the last three years. We select some important papers. Reference\cite{21_duan2023distributed} takes EC as the bridge connecting CC and IoT layer by describing EC's fundamental features. Reference \cite{23_salaht2020overview} concentrates on the Service Placement Problem (SPP) in Fog/EC, discussing classification, advantages, disadvantages, taxonomy, and optimization strategies. They dive into SPP of a control plan and optimization strategies. Reference \cite{24_hua2023edge} defines EC, discusses the combination of EC and AI, and covers challenges such as task offloading, resource allocation, and security. They claim EC's shortages could be fixed with the help of machine learning techniques and give their different applications. Reference \cite{25_vhora2020comprehensive} provides a brief introduction to the related concepts like Fog computing, cloudlets, and MEC and discusses EC challenges. Their research covers different EC simulation tools. Reference \cite{26_chang2021survey} summarizes EC architecture, keyword technologies, security, privacy protection, and applications. They list comparatively hot research zones in EC, like resource allocation, offloading, mobility management, etc. They also cover common security issues, including data security sharing, authentication in a domain, etc.  In addition, they review 5G as well as commonly used software and hardware systems for EC. Reference \cite{28_khan2020edge} analyzes EC paradigms' evolution and provides a comprehensive taxonomy for smart cities. They give detailed comparisons among CC, Cloudlets, Fog, and MEC. In the field of smart cities, their taxonomy includes security, edge analytics, edge intelligence, resource, caching, resource management, characteristics, and sustainability. In addition, they also give two practical cases and future research directions in the field of smart city. Reference \cite{29_filali2020multi} illustrates the integration of MEC, 5G, and mobile networks and focuses on MEC resource management. Driven by 5G, they focus on QoS because it is the critical factor to cope with latency in EC. They also list detailed techniques for the integration, including FV, SDN, and SFC.  Reference \cite{30_luo2021resource} discusses different collaborative methods for resource scheduling under EC. They cover several terms used in resource scheduling in EC, including resources, tasks, participants, objectives, actions, and methodology in their research model. 

Summary: This section summarizes the existing surveys covering the topic of EC. They include different EC's concepts, architecture and paradigms, resource allocation, and specific optimization problems including scheduling. The resource allocation in EC is inevitably the kernel and hot topic under different EC circumstances.

\subsubsection{IoT-EC-oriented Surveys}

In Table I, we have included many surveys that cover the IoT on EC paradigm. In reference \cite{23_salaht2020overview} IoT's requirement is also taken as the driving factor for EC development. The IoT deployment in Fog layers and services among IoT and Fog/Cloud layers. In \cite{24_hua2023edge}, the authors explain the reason for employing EC for IoT. In their research, the IoT requirements trigger the emergence of EC as the solution. They cite urban healthcare utilizing IoT. Reference \cite{25_vhora2020comprehensive}  covers IoT architectures and requirements. They also list several simulation tools for both EC and IoT by implementing different environments. Reference \cite{29_filali2020multi} focuses on IoT requirements as the driving factor in the integration of 5G and MEC. They do not dive into the details of IoT in their research. The authors of \cite{30_luo2021resource} connect IoT and EC in IoT-EC collaboration and IoT-EC-CC collaboration, stressing IoT’s role in different diagrams, from the perspective of different collaboration methods in resource allocation. They also list several resource allocation methods in the context of IoT-EC applications. 

Summary: This section further connects EC with the IoT. Previously, EC research is performed in cooperation with the MEC circumstances. During their research, more focuses are shifted to the limitation of IoT devices, eg the battery lifespan, resource limitation, and the optimal strategy to fix these constraints. Only one paper covers a compromising domain of simulation tool.

\subsubsection{AI-oriented Surveys} 
Reference \cite{21_duan2023distributed} focuses on distributed learning algorithms including synchronous SGD, asynchronous SGD, and gossip SGD. They also list several distributed learning frameworks, and optimization technologies for distributed learning.  Reference \cite{121_deng2020edge} gives two taxonomies pertaining to AI and EC, 1) AI for edge to optimize edge computing by determining policy of resources allocation, and 2) AI execution on edge servers covering training and inference phases. Some surveys primarily focus on AI in the context of IoT and EC, including the enhancing security of authentication\cite{24_hua2023edge}\cite{30_luo2021resource}. These surveys concentrate on optimizing resource management, latency, energy consumption, and security. Reference \cite{24_hua2023edge} emphasizes traditional ML and DL, exploring their different models and optimization in EC scenarios such as computing offloading, energy consumption reduction, security, and resource allocation. They also cover the applications of machine learning in EC under the circumstances of smart city, smart manufacturing, and the IoV.  Reference \cite{30_luo2021resource} covers machine learning as a solution for complex optimization problems, including reinforcement and deep learning for optimal decisions in dynamic environments. It highlights machine learning's strengths and weaknesses in different circumstances including centralized and distributed methods of EC. 

Summary: This section compared different AI research topics in existing surveys and reviews. By beginning the EC requirements (offloading, latency), they also acknowledge the limited capacity of EC and lead to the dual direction benefits by cooperating machine learning with EC.

\subsubsection{This Survey}

While the existing surveys and reviews cover IoT, EC, and CC, only \cite{24_hua2023edge} and \cite{30_luo2021resource} delve into AI. To the best of our knowledge, our survey is the first to comprehensively consider IoT, EC, CC, and AI from a partitioning perspective.  We finally provide insights into future research directions.

\begin{table*}[htbp]
\small
\begin{center}
\caption{A Taxonomy of Partitioning in IoT-Edge-AI Environment}

\begin{tabular}{ l l c c c c c c c c c c }
    \toprule
    \textbf{Ref} & \textbf{Description} & \textbf{IoT} &\textbf{EC}  & \textbf{CC} & \textbf{AI}  & \textbf{Open}  & 
      \multicolumn{3}{c}{\textbf{Partitioning Technique}}\\
      \cmidrule(r){8-10}
     &  & & & &  & \textbf{Challenges} & \textbf{Data} & \textbf{Task} & \textbf{DNN Model} \\
     &&&&&&& \textbf{Partitioning} & \textbf{Partitioning} &\textbf{Partitioning}\\ \hline

            \cite{21_duan2023distributed}   & End-Edge-Cloud       & \checkmark & \checkmark & \checkmark & \checkmark & \checkmark &            &  & \checkmark &     \\ \hline
           \cite{23_salaht2020overview}    & Service Placement     & \checkmark & \checkmark & \checkmark &            &            & \checkmark & \checkmark & \checkmark &     \\ \hline
           \cite{24_hua2023edge}           & Edge AI               & \checkmark & \checkmark & \checkmark & \checkmark & \checkmark &            & \checkmark & \checkmark &     \\ \hline
           \cite{25_vhora2020comprehensive}& Mobile Edge           & \checkmark & \checkmark & \checkmark &            & \checkmark &            &            & \checkmark &     \\ \hline
           \cite{26_chang2021survey}       & Edge research         & \checkmark & \checkmark & \checkmark &            & \checkmark & \checkmark & \checkmark & \checkmark &    \\ \hline
           \cite{27_cao2020overview}        & Recent Advances       & \checkmark & \checkmark & \checkmark & \checkmark &           & \checkmark & \checkmark & \checkmark &
           \\ \hline
           \cite{28_khan2020edge}          & Smarty City           & \checkmark & \checkmark & \checkmark &            & \checkmark & \checkmark &            & \checkmark &    \\ \hline
           \cite{29_filali2020multi}       & MEC                   & \checkmark & \checkmark & \checkmark &            &            &   & \checkmark          & \checkmark &    \\ \hline
           \cite{30_luo2021resource}       & Resource Scheduling   & \checkmark & \checkmark & \checkmark & \checkmark & \checkmark &   & \checkmark          & \checkmark &    \\ \hline
           
            \makecell{This \\Survey}       &                       & \checkmark & \checkmark & \checkmark & \checkmark & \checkmark & \checkmark & \checkmark & \checkmark &    \\ 
    \bottomrule
  \end{tabular}

\end{center}
\end{table*}

\subsection{Contributions of this survey}

In this survey, our contributions are outlined as follows: 
\begin{itemize}

\item We provide a comprehensive analysis of the synergy among IoT, EC, CC, and AI from the partitioning perspective. We highlight the taxonomy, new requirements, and the advantages of the partitioning techniques. Additionally, we delve into different EC paradigms, ML, DNN, and IoT for a better understanding of partitioning.

\item We discuss partitioning techniques under three categories: data partitioning, computation task partitioning, and DNN model partitioning, in IoT-Edge-AI. We also list several potentially useful and relevant research directions for partitioning techniques. We stress the requirements and motivations of each partitioning technique. To the best of our knowledge, our survey is the first to approach this topic from the partitioning perspective.

\item We present the need for the integration of different partitioning techniques among IoT, Edge, and AI for future Internet applications.

\end{itemize}

\subsection{Survey Organization}
This survey is organized as follows: Section 2 explores the background of CC, EC, IoT, and AI. Section 3 discusses applications resulting from the combination of EC, IoT, and AI. Section 4 serves as the core of this survey, analyzing the taxonomy of different partitioning techniques.  Section 5 addresses future research in this combined domain. Finally, Section 6 provides a summary and conclusion for the survey.

\section{CONCEPTUAL BACKGROUND}
This section contains conceptual information that should help bring proper context and perspective to the discussion that follow. We believe that this will set-up appropriate background and features of CC, IoT, EC, and AI,  for later discussing the use of partitioning.

\subsection{Cloud Computing Architecture: its Advantages and Disadvantages}

Since virtual computers became popular in the late 1990s, Amazon has employed CC to handle the gap between peak and trough requirements during events like Black Friday. Today, CC, a widely discussed topic in academia and industry, draws inspiration from the concepts of distributed software architecture and virtualization. Its primary goal is to provide infinitely elastic, hosted services over the internet. Based on its capability of elastically delivering services, CC has given rise to various applications, such as data storage, big data analysis, E-commerce, and more, from centralized data centers strategically located across diverse regions globally. There are three main traditional types of CC models one new type of CC model \cite{32_alouffi2021systematic}: The traditional ones are Software as a Service (SaaS), Platform as a Service (PaaS), and Infrastructure as a Service (IaaS). Container as a Service emerges as a new cloud model aimed at dissociating applications from PaaS environment specifications. There are several types of CC forms, including private clouds, public clouds, hybrid clouds, and multi-clouds, that have been put into practice. Private clouds are tailored for a single organization, offering enhanced security and control over data. On the other hand, public clouds are owned and operated by third-party service providers, delivering scalability and cost-effectiveness but with some trade-offs in control. Hybrid clouds combine private and public cloud resources, allowing data and applications to move seamlessly between them. Multi-cloud environments involve using multiple cloud providers for different services, providing redundancy and optimization, but also requiring robust management and integration strategies. These cloud models offer organizations a spectrum of options to meet their specific needs, balancing factors like security, performance, and cost efficiency.

In practice, CC offers several advantages, including 1) trading fixed expenses for variable expenses based on end users’ resource usage; 2) benefiting from massive economies of scale through virtualization technology; 3) eliminating the need to guess capacity as CC can adjust resources elastically; 4) increasing speed and agility with a powerful yet user-friendly dashboard; 5) reducing costs associated with running and maintaining data centers; and 6) enabling global reach within minutes through the services of CC providers \cite{aws_6}.

Despite being empowered with high-performance hardware platforms and proven to be suitable for many applications, CC has some drawbacks, particularly when handling time-sensitive services : 

1) Latency: While CC providers make efforts to distribute their data centers as widely as possible; there is inevitably a significant distance from end-users and end-equipment to these centers. On the other hand, many applications do not demand computation capacity as high as CC platforms but require time-sensitive inference services. For example, time-sensitive robotic services in E-healthcare require a latency of no more than 300 milliseconds round-trip time. Uploading data to the cloud remotely for inference may encounter unacceptably high interval delays in the network and thus fail to meet the requirement of end-to-end low latency, which is known as data access latency \cite{33_tang2022latency}. Access latency hinders the widespread adoption of CC in IoT applications.

2) Privacy: CC typically involves storing the complete dataset at data centers. Sending sensitive data to the cloud can raise privacy concerns among users. For instance, there are concerns related to patient information confidentiality, privacy, and service costs \cite{18_mehrtak2021security}. In addition, there's the potential for the cloud to misuse data for research purposes, further eroding trust between the cloud platform and end-users.

\subsection{Different Edge Computing Architectures}

Differing from traditional cloud-centric computing, which runs on cloud servers, EC is relatively a new computing paradigm that performs computations, data storage and network control at the edge of the network \cite{34_garg2021security}. Its primary goals are: 1) reducing the traffic that travels over the Internet; 2) reducing the latency of long-haul communications between end-users and traditional CC servers \cite{35_carvalho2021edge}. By 2023, there have been several EC-related architectures proposed, like the following: 1) Multi-access EC; 2) Cloudlet Computing; and 3) Fog Computing;.

\subsubsection{Multi-access Edge Computing (MEC)} 

Multi-access Edge Computing (MEC), previously named Mobile Edge Computing before 2016, has evolved to signify and deploy computation tasks at the network's edge. MEC is designed to establish a seamless and open framework, facilitating the integration of diverse EC applications, including network, data storage and processing centers, and computation providers \cite{36_ranaweera2021survey}. The European Telecommunication Standards Institute (ETSI) has outlined five key characteristics of MEC, including on-premises, allowing MEC platforms to operate independently from the broader network while providing proximity, ultra-low latency, and high bandwidth \cite{37_}. In essence, MEC serves as a standard for mobile communications.

\subsubsection{Cloudlet Computing} 

In 2009, the concept of cloudlets was first proposed, serving as a precursor to Edge and Fog Computing. Cloudlet aims to provide flexibility, mobility, scalability, and elasticity. They enable data centers to serve mobile or IoT devices within single-hop proximity through a high-bandwidth, low-latency network. Virtual Machine (VM) is the foundation of Cloudlet and allows independent operations from cloud services \cite{35_carvalho2021edge}. Cloudlet also owns a three-layer architecture: 1) the mobile devices layer, 2) the micro cloud layer, and 3) the cloud layer.

\subsubsection{Fog Computing} 

In 2012, Cisco introduced Fog Computing for the proliferation of IoT applications with requirements of time-sensitive data processing. Fog computing primarily focuses on processing and storing data at the edge side. In contrast to Cloudlets and MEC, Fog Computing places more emphasis on the IoT and end-side processing \cite{38_sabireen2021review}. The architecture of Fog computing owns three layers: 1) CC layer, 2) Fog layer, and 3) terminal layer.

\subsection{IoT concept,  Structure, Protocols and Requirements}

Though there is no universally accepted definition for IoT as of now, it can be defined as “The Internet of Things (IoT) refers to a network of physical devices, vehicles, appliances, and other physical objects that are embedded with sensors, software, and network connectivity, allowing them to collect and share data.” \cite{105_dorsemaine2015internet}. A typical IoT architecture is depicted in Figure 2. This architecture integrates cyber and physical components, harnessing modern sensing, computing, and communication technologies \cite{116_ang2022towards}. IoT aims to create a comprehensive, perceptual, and universally connected environment for end-users through networks and sensors. In Figure 2, there are three layers in IoT architecture: 1) the perception layer, 2) the network layer, and 3) the application layer. We describe each layer as follows:
\begin{figure}[!t]
  \centering
    \includegraphics[width=0.5\textwidth]{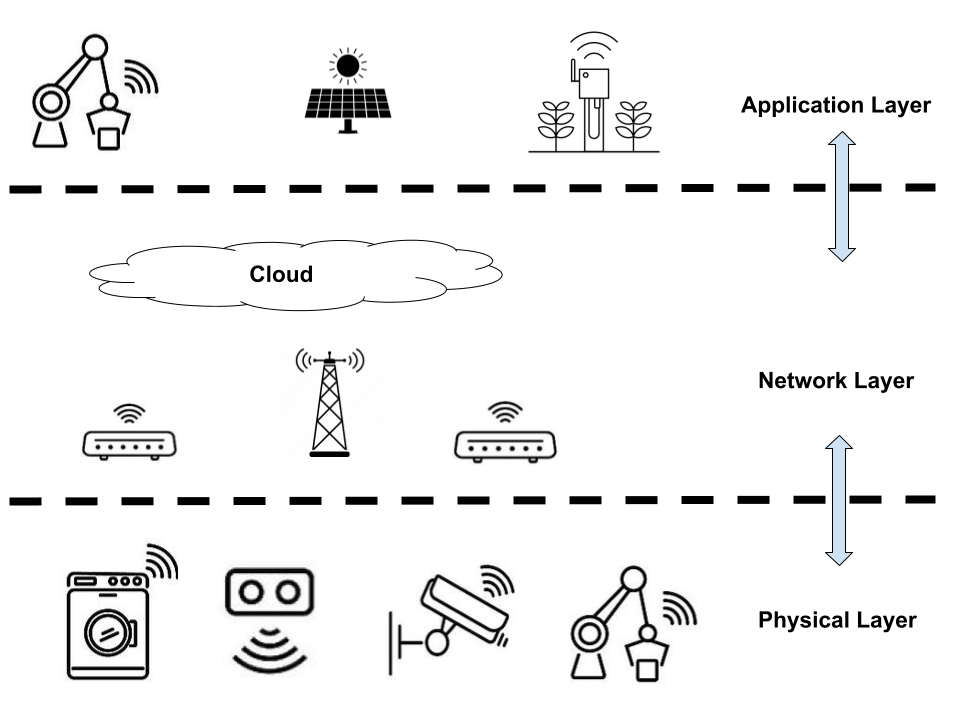}
    \caption{An architecture of the IoT}
    \label{fig:2}
\end{figure}

1) Physical layer or Perception Layer: Generally, this layer consists of wireless sensors, batteries, and actuators. It serves as the sensor layer of the IoT structure, allowing IoT devices to collect real-time data from the surrounding environment through distributed sensors together with technologies like actuators, radio-frequency identification (RFID) devices, 2-D codes, batteries, and more. The collected data may include different attributes, such as location information, audio data, video data, or environmental response data.

2) Network Layer: The network layer is the backbone of IoT and is used for data transmission. IoT devices communicate with gateways, Wi-Fi networks, or base stations to securely transfer data from the perception layer. This layer incorporates various network protocols designed for both short-range and long-range transmissions.

3) Application Layer: The application layer functions as the interface through which end-users access the necessary services or operations. This layer comes into play after a vast amount of polymorphic and heterogeneous data is processed and analyzed. In Figure 2, examples of applications in this layer include but are not limited to industrial control, traffic management, inventory management, and green agriculture.

Traditionally, IoT devices connect to the cloud through nearby edge gateways using various data protocols through short-distance protocols of Bluetooth, Wi-Fi, ZigBee, and 6LoPWAN,  or long-range protocols like NB-IoT, LoRaWAN, and 5G\cite{1_kong2022edge}.  All these abbreviations are listed and explained in Table VIII in the Appendix.

Different from traditional PC devices, IoT devices own limited and heterogeneous computation resources and storage resources. Correspondingly, IoT devices always produce imbalanced and non-iid data from different contexts. However, IoT applications have high QoS requirements with low latency and the capability to process heterogeneous types of data. On the other hand, IoT system designers are concerned with the system scalability in practice. 

\subsection{AI Algorithms, Framework in an IoT-EC-CC Environment}

AI models, specifically deep learning modes, are inherently designed to analyze, learn from and produce actionable insight from large amounts of data. Researchers have successfully applied AI to handle large amounts of data generated by the multitude of heterogeneous devices that IoT applications use. Some of the fields in which AI has found application include healthcare and autonomous vehicles, achieving remarkable performance. In many cases, Machine Learning, a subset of AI, plays a dominant role in learning patterns from data and making automated decisions based on these patterns. In summary, AI is most used in data analysis and allocating resources in the existing studies.  Deep Learning gained significant attention in the field. Additionally, Reinforcement Learning has shown promise in solving decision-making problems. More recently, a new method known as Deep Reinforcement Learning has been developed to enable learning from the environment, with AlphaGo being a notable success. 

In this section, we will review traditional Machine Learning algorithms, Deep Learning algorithms, and Deep Reinforcement Learning methods and touch upon topics like federated learning and evolutionary algorithms. We list several commonly used algorithms and frameworks related to the field of IoT-EC-CC. 

\subsubsection{Traditional Machine Learning Algorithms} 

Machine learning is one of the most critical fields in the modern computing application market. Machine learning can learn from training data by analyzing and extracting the model and finally solving relevant tasks automatically. Traditional machine learning algorithms are classifiable into three categories: supervised learning, semi-supervised learning, and unsupervised learning, according to the accessibility of label information,
Although they have been employed in many scenarios with considerable accuracy, traditional ML algorithms have limitations such as a lack of interpretability, bias, dependency on data quality, data privacy, security concerns, and high false positives. However, they are generally easier to deploy and require fewer computational resources in practice compared to DL and RL.

\subsubsection{Deep Learning (DL) Algorithms} 

DL represents an evolution from the conventional ML algorithm, specifically Neural Networks, characterized by an extensive number of hidden layers. DL can derive high-level features from raw data by computer, eliminating the necessity for feature engineering typically associated with traditional ML models. In practice, DL is generally equivalent to DNN. In comparison to traditional ML algorithms, DL excels at extracting high-level features from raw data, albeit at a higher computational complexity and data volume, often driven by the availability of high-performance GPUs. Commonly used Deep Learning models and corresponding examples are listed in Table II. During the execution of Deep Learning models, the data transferred between layers generally decreases, making it possible to transmit less data and reduce network workload when the layers are distributed across the EC architecture. DL excels in various fields such as generative AI, computer vision, and autonomous driving, achieving higher accuracy. Additionally, DL is applied in EC studies for tasks like task offloading strategy, resource allocation, optimization in energy, and latency \cite{24_hua2023edge}.  Table II, nine commonly used DL algorithms are listed since DNN models are the kernel of this survey. 

Different from traditional ML algorithms, DL excels in various fields such as generative AI, computer vision, and autonomous driving, achieving higher accuracy. Additionally, DL is applied in EC studies for tasks like task offloading strategy, resource allocation, optimization in energy, and latency \cite{24_hua2023edge}.

\subsubsection{Reinforcement Learning (RL), Deep Q Network (DQN) and Deep Reinforcement Learning (DRL)} 

Both supervised learning and unsupervised learning depend on predefined datasets to predict new data based on the patterns they have learned. In contrast, RL constructs models based on the dynamic interactions with the external environment. Based on historical experience, RL collects the state of the environment and performs actions to maximize the reward, which serves as the criteria for the outcome. Typical RL algorithms include the value-based Q-learning algorithm, which is based on an expected reward known as the Q-value updated in each iteration. DQN is a DNN-based approach used to approximate the Q-values, to achieve optimal decisions based on current and historical rewards. DRL combines DL and RL, where DL approximates the Q-value, and RL offers the decision-making ability, which evaluates expected outcomes by mapping the current state to corresponding actions. RL excels at solving decision-making problems and finds applications in EC resource management and task offloading allocation \cite{40_alfakih2020task}. DLR, also valuable for decision-making, enhances the efficiency of network or computing resource allocation policies\cite{31_liu2020resource}. 

\subsubsection{Federated Learning (FL)} 

In 2017, FL emerged as a distributed Machine Learning framework, initially developed by Google to take care of privacy concerns. In many scenarios, users are sensitive to uploading their private data to centralized CC servers for processing. FL eliminates the need for different local users to upload all their data to the server. Instead, they train a local model using their data respectively and then aggregate them into a global model on a central server. This approach allows for the partitioning of the global model into local models.

\begin{table*}[htbp]
\small
\caption{Examples of DL Algorithm and their Applications}

\begin{tabular}{ |p{0.35\textwidth}|p{0.5\textwidth}|p{0.05\textwidth}| } 
 \hline
\textbf{DL Model Examples}	& \textbf{Applications Examples}	 & \textbf{Ref}\\
\hline
Convolutional Neural Networks (CNNs)	& Image classification, human activity classification 	&  \cite{107_zhu2021lightweight}  \\
\hline

Long Short Term Memory Networks (LSTMs) & Speech recognition, music composition	& \cite{108_wu2021edgelstm}  \\
\hline
Generative Adversarial Networks (GANs)	& Game development of 2D resolution	&\cite{110_kim2022deep}\\
\hline
Recurrent Neural Networks (RNNs)	    & Spatiotemporal predictive learning 	& \cite{111_wang2022predrnn}\\
\hline
Multilayer Perceptrons (MLPs)	        & Classification of diabetes	& \cite{112_sivasankari2022classification}\\
\hline
Self-Organizing Maps (SOMs)             & Identify the spacing and position of high dimensional clusters	& \cite{113_hasan2022edge}\\
\hline
Deep Belief Networks (DBNs)             & Motion capture data, speech recognition	& \cite{115_almogren2020intrusion}\\
\hline
Restricted Boltzmann Machines(RBMs)  	& Security and optimal service composition for workflows	& \cite{118_lakhan2022restricted}\\
\hline

\end{tabular}

\end{table*}

\subsection{Evolutionary Algorithms}

Evolutionary computation is an approach inspired by the theory of Darwinian evolution, rather than following first principles and executed in a stochastic manner. It furnishes effective approximate solutions for problems that are challenging to address, particularly in cases where fundamental principles are lacking but a near-optimal solution proves satisfactory. Evolutionary algorithms typically begin by initializing variables and then simulating the evolution process, which includes selection, population reproduction, variation, and population updating in each iteration. In EC, many problems including resource optimization\cite{44_an2022joint}, workload balancing \cite{43_kuang2020offloading}, and task scheduling\cite{45_al2020task},  utilize Evolutionary Algorithms.

\subsection{Summary}
In summary, conclude with what we learned from the background discussion. This section provides a necessary and detailed background covering EC, CC, IoT, and AI. It extends the Introduction section from a historical and requirements perspective and paves the foundation for Section 3.

\section{PARTITIONING IN IoT-EC-CC-AI}

Having discussed the architecture, and background technologies in IoT, EC, CC, and AI, we are ready to discuss the role of partitioning in this environment. This section is the kernel section of this survey covering the partitioning techniques. In IoT-EC-AI, our partitioning technique is discussed in three dimensions: 1) Data Partitioning; 2) Computation Task Partitioning; and 3) DNN Model Partitioning. 

\subsection{Data Partitioning}

In this survey, data partitioning is defined as dividing a dataset into sub-datasets for distributed processing learning like complexity reduction and reduced latency. Data partitioning has been successfully employed in MapReduce for Big Data Analytics. Traditionally, there are three types of data partitioning: vertical partitioning, horizontal partitioning, and sharding \cite{98_charlie}. Vertical partitioning moves some columns into separate tables with the same number of rows but different schemas. Horizontal partitioning divides data into multiple smaller tables with the same schema. Sharding is also a horizontal partitioning technique but among different servers. In the context of the substantial data output from IoT devices, EC introduces a myriad of challenges, encompassing issues related to security, data incompleteness, and significant upfront as well as ongoing expenditures \cite{49_ganesh2022improving}. Naturally, the data is generated among IoT devices and then uploaded to EC servers. At centralized CC servers, the machine learning model stores and uses the entire training dataset locally. In \cite{17_ale2022d3pg}, they proposed an algorithm called Dirichlet deep deterministic policy gradient to optimize task offloading with task partitioning and computation resource allocation. In their design, they propose a reply buffer with a fixed size to hold updated data in sequence in the training phase. Data partitioning is capable of making applications more scalable and eliminating the security challenges of uploading the complete dataset.

In Table III, we collect papers mentioning data partitioning for security purposes. It lists where data are spread in the EC-CC context and what security method is used. The data partitioned falls into several categories: on IoT devices, on EC servers, and on CC servers. We categorize these studies into different implementation methods focused on security:

\begin{table*}[htbp]
\small
\centering
\caption{ Works that Discuss Data Partitioning for Security}

\begin{center}

\begin{tabular}{ |p{0.03\textwidth}|p{0.3\textwidth}|p{0.18\textwidth}|p{0.25\textwidth}| } 
 \hline
	\textbf{Ref} & \textbf{Description} &\textbf{Data location} & 		\textbf{Method Used}  \\
 \hline
\cite{99_sun2020toward}  &Communication-efficient federated learning &EC server	&  Federated Learning \\
\hline
\cite{100_bao2021edge} & Edge computing-based &EC server &	 	Federated Learning  \\
\hline
\cite{47_tianqing2021resource} & Resource allocation &EC server	&	Federated Learning \\
\hline
\cite{22_khan2022federated} & Federated Split & EC server & Federated Learning \\
\hline
\cite{50_duan2022combined} & Split learning in & EC server & Split Learning \\
\hline
\cite{52_kim2023bargaining} & Bargaining Game & EC server & Split Learning \\
\hline
\cite{67_ismail2023analyzing} & SplitFedLearning & EC server & \makecell[l]{Hybrid of Federated \\Learning and Split Learning} \\
\hline
\cite{69_shen2023ringsfl} & Ringsfl & EC server & \makecell[l]{Hybrid of Federated \\Learning and Split Learning} \\
\hline
\cite{75_gao2020end} & End-to-End & EC server & \makecell[l]{Hybrid of Federated \\Learning and Split Learning} \\
\hline

\cite{48_prabhu2022edge}& Healthcare framework & EC server	&  APIs	 \\
\hline
\cite{49_ganesh2022improving}& Security in Edge Computing & EC server, CC server	& Rating Table	  \\
\hline

\cite{51_gao2022ppo2}& Location privacy-oriented &EC server	& PPO2	 \\
\hline
\cite{53_dong2022integration}& Integration of edge & EC server, IoT devices &	 Fusion Method	 \\
\hline

\cite{54_li2020secured}& SDN-based edge &EC server	& 	Authentication scheme \\
\cite{46_cheng2021blockchain} & Blockchain-Based & EC server & Authentication scheme \\
\hline

 \hline
\end{tabular}

\end{center}

\end{table*}

\subsubsection{Federated Learning Framework}
As mentioned earlier, Federated Learning is driven by privacy concerns related to data storage, and it exploits partial data availability at different nodes. Federated learning allows data to be trained on distributed edge devices and then merged into a complete model, avoiding the need for the global server to access the complete data and thereby protecting privacy. In addition, in some studies, data is partitioned and processed at different locations for security but not through the Federated Learning algorithm. In Figure 4, the framework is depicted in steps. In step 0, the main server initializes and distributes a unique model with hyper-parameters and training rate to all EC servers. In each epoch, 1) EC servers train the model locally with their private data; 2) the model parameters are delivered to the main server to aggregate; and 3) the updated model parameters are sent to EC servers for the next epoch. From the framework description, FL obviously has a high cost of communication between the main server and EC servers. 

\begin{figure}[!t]

    \includegraphics[width=0.5\textwidth]{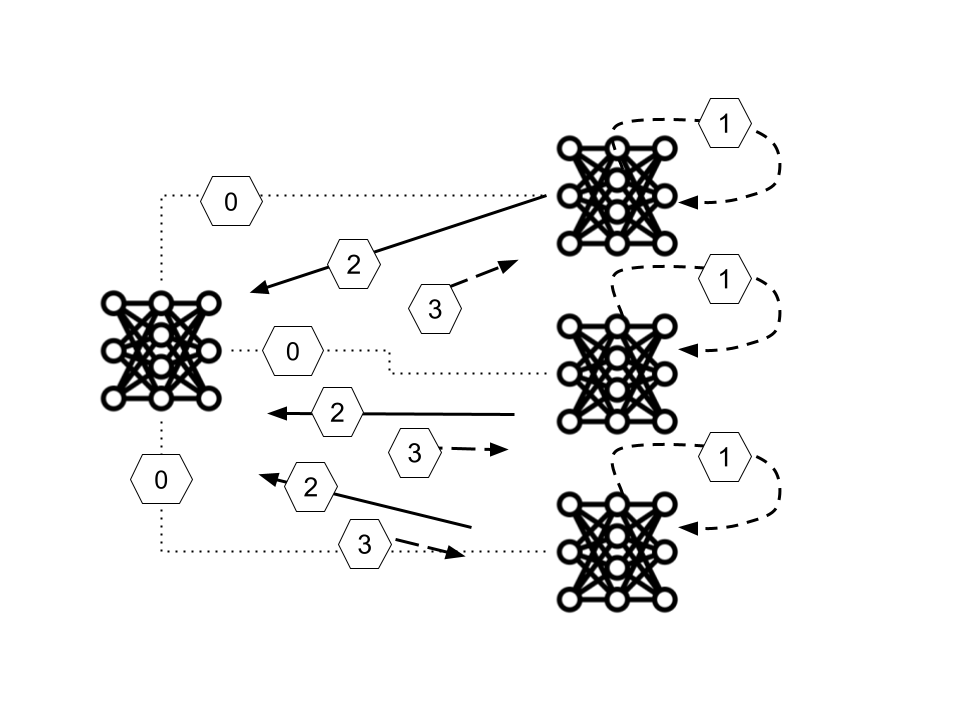}
    \caption{The framework of Federated Learning}
    \label{fig:3}
\end{figure}

The Federated Learning (FL) framework has been utilized in a sequence of studies. In \cite{47_tianqing2021resource}, the authors propose a schema combining federated learning, reinforcement learning, and concurrent decision-making. Their algorithm ensures both global optimality and better privacy guarantees than traditional federated learning. Reference \cite{99_sun2020toward} proposes a general gradient sparsification (GGS) framework comprising 1) a gradient correction optimizer to make the model converge better by properly treating the accumulated insignificant gradients; and 2) a batch normalization (BN) update local gradients to mitigate the delayed gradients. Their design does not increase the burden of data transmission. Their experiment shows that GGS achieves the accuracy with 99.9\% of gradients sparsified. Reference \cite{100_bao2021edge} focuses on the context of autonomous driving and uses edge vehicles (some selected vehicles) as FL clients to train the local models. Their model achieves high accuracy but also makes a tradeoff between accuracy and communication overhead. Besides the data privacy, in Ref \cite{22_khan2022federated}, since data poisoning in the later training phase learns inaccurate features among input and labels, they proposed DepoisoningFSL as the data poisoning mechanism in EC, obtaining high accuracy and low error rate. They use random forest (RF), k-nearest neighbor (KNN), and linear regression(LR) to identify correct labels. 

\subsubsection{Split Learning}
Split Learning (SL) constitutes a collaborative learning paradigm wherein a Machine Learning (ML) model undergoes segmentation into distinct components, each trained independently yet collaboratively. In the context of Internet of Things (IoT) devices, the distributed execution of segmented models transpires locally, utilizing device-specific data, with subsequent transmission of intermediate outcomes to a server. Figure 5 depicts a Split Learning framework. Different from the FL which trains local models at EC servers in parallel, the SL framework with multiple EC servers uses data locally and trains different models, for instance, layers of DNNS,  in a round-robin style. All EC servers take turns to train local models with alternating epochs in cooperation with the main server. In Figure 4, at the beginning, the complete model is split into 4 units, one running at the main server(could also be EC server) and the other three running at EC servers. In turn, each EC server forwards local propagation until the last activation layer of the model assigned to calculate smashed data in order. Then, the smashed data is uploaded to the main server which completes its forward propagation. Then the main server performs the back-propagation which computes its gradients and its smashed data,  and then sends it back to the EC servers for their own back-propagation. By now, a single pass of back-propagation is completed and continues the next pass till all EC servers reach a decent convergence. 
In the conventional SL framework, numerous EC servers engage in collaboration either in a Peer-To-Peer model or centralized training with a singular server, collectively training a global model by harnessing the private data stored on diverse user devices. This approach also enables user devices to delegate a fraction of the model training responsibility to a server, thereby facilitating the exploitation of flexible resource management within edge computing infrastructures to support ML model training. While SL and FL entail local model training via private data with subsequent aggregation at a central server, the key distinction lies in FL's pursuit of a unified model training across all clients, whereas SL entails the training of distinct partitioned models. FL has drawn attention in studies recently, comparatively,  there are much fewer studies on SL\cite{50_duan2022combined}. In Ref \cite{52_kim2023bargaining}, they propose a Spling Learning framework that allows the configuration of cut layers to achieve optimization in balancing energy consumption and training time. They take the balancing problem as a multiplayer bargaining problem, solved by the Kalai-Smorodinsky bargaining solution.

\begin{figure}[!t]
  \centering
    \includegraphics[width=0.5\textwidth]{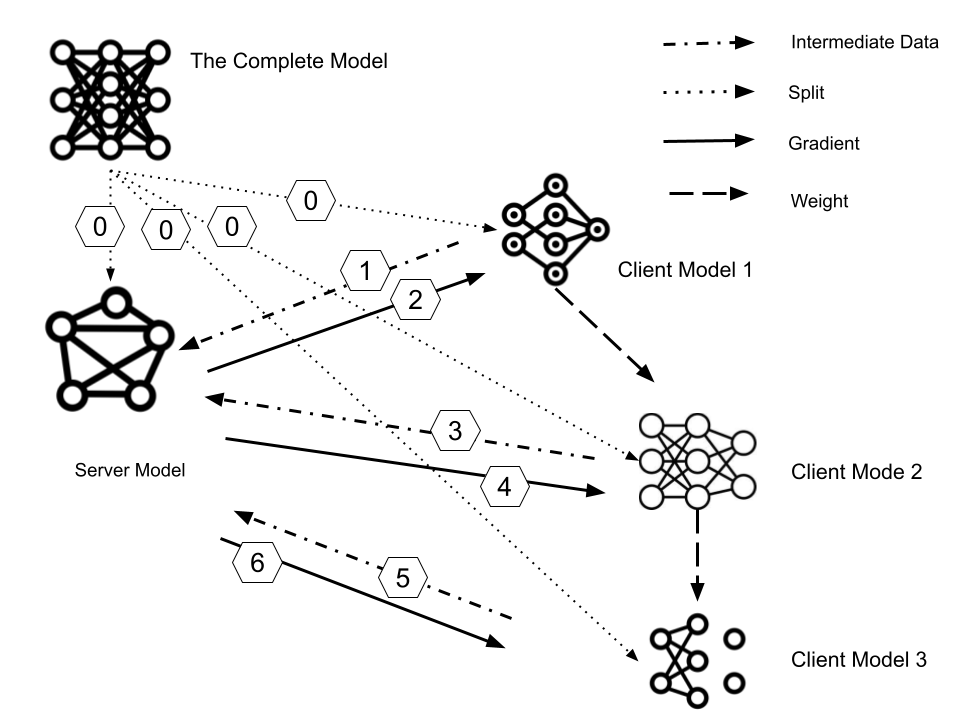}
    \caption{The framework of Split Learning}
    \label{fig:4}
\end{figure}

\subsubsection{Hybrid of Federate Learning and Split Learning}
Though FL performs better than SL in communication cost, neither is suitable for heavy models trapped by the too-long training time\cite{75_gao2020end}. They also compare the performance FL and SL under imbalanced or non-IID datasets. They claim SL exceeds FL in an imbalanced dataset, while FL surmounts SL in extreme non-IID dataset. In addition, both FL and SL have their internal shortages. In Ref \cite{67_ismail2023analyzing}, they claim only one client at an instance is training as other clients are idle in SL, leading to longer training time. While the FL can train clients in parallel, they propose the hybrid of FL and SL,  namely SplitFed Learning (SFL) of Distributed Collaborative Machine Learning, to fix the problem. They evaluate it by analyzing the impact of data poisoning attacks to evade the SFL aggregated model.  In Ref \cite{69_shen2023ringsfl}, they claim FL suffers from the heterogeneity of computational capability and battery level among the clients involved in the training phase. For instance, the uneven computational resources and battery level of clients lead to divergent training times respectively, resulting in the straggler effect and undermining the FL efficiency. They propose the RingSFL by integrating FL and SL, where neighbor clients can communicate in wireless link. The server assigns a propagation length to each client to constrain the traversing of the ring topology in the forward and backward propagation to complete the training. RingSFL can empower data privacy, utilize computational resources, and allocate workload according to the system heterogeneity. 

\subsubsection{Using Authenciation}

In addition to federated learning, authentication can also be used for data security. Reference \cite{54_li2020secured} proposed a framework with an SDN controller to balance and optimize resource utilization. They use a lightweight authentication scheme and Edge servers to store and process data from the patients. In Ref \cite{46_cheng2021blockchain}, they claim since bogus EC computation nodes pose threats to IoT-EC applications, a blockchain-based mutual authentication scheme driven by certificateless cryptography is proposed for data privacy.  

\subsubsection{Using Fusion method}
Reference \cite{53_dong2022integration} provides a lightweight data fusion method to secure data in IoT by executing the training process at EC nodes locally. Through fusion methods, no data but the trained model is uploaded to CC servers.

\subsubsection{Using APIs}
The primary intent of employing a REST API resides in the simplification of intricate business models, facilitating streamlined communication and data exchange. Within this framework, task distribution ensues as clients submit requests to the API, seeking execution of designated actions or retrieval of specific information. Naturally, REST API can be employed for data partitioning by distributing datasets to different servers. In \cite{48_prabhu2022edge}, the Device Service first initializes a sensing device. Then, the sensing device initiates the transmission of data. The Device Service, responsible for orchestrating these operations, undertakes the translation of sensor data into an EdgeX event object. The Core data service receives this event object through REST communication. Finally, they keep the transmitted data in RedisDB, denoting a local edge database. 

\subsubsection{Using Rating Table}
Caching data can also be used to protect data in data partitioning. In \cite{49_ganesh2022improving}, their architectural design integrates data centers and edge servers, each featuring a rating table to meticulously record information about all devices. This table operates by caching content retrieved from CC resources, wherein these devices have submitted ratings to mutually benefit each other. 

\subsubsection{Using PPO2}
In \cite{51_gao2022ppo2}, they assume data location is sensitive information, and data is distributed and marked by different vehicles. They claim that attackers with a
priori knowledge can infer the user’s location information based on the user offloading event. To fix this risk, they propose PPO2 method to both plan offloading tasks and reduce attack loss.    

\subsection{Computation Task Partitioning}

In the IoT-EC-CC environment, data and computation from IoT have to always be offloaded to EC-CC, due to the limited resource capacity of IoT devices. In CC, almost all computation tasks are simply uploaded to the cloud servers to complete all processing jobs by utilizing their high-performance hardware platform. In EC, however, given that EC servers usually own limited computation resources and are limited by network resources, computation task offloading technology offers a feasible solution by distributing some computation-intensive tasks to the CC servers, EC servers, or hybrid. Partitioning computation tasks under different constraints becomes challenging when designing and employing EC applications due to two factors: 1) adaptability, computation tasks need to be distributed among EC nodes dynamically, and 2) effectiveness, offloading the worth tasks to make the best use of system resources.

In Table IV, we summarize papers focusing on computation task partitioning algorithms and frameworks on EC or IoT. We categorize them by their different goals, 1) security, 2) QoS(low latency, better network bandwidth, accuracy).  

\begin{table*}[htbp]
\small
\centering
\caption{Paper of Computation Task Partitioning Related Algorithm/Framework}

\begin{center}
\begin{tabular}{ |p{0.03\textwidth}|p{0.1\textwidth}|p{0.2\textwidth}|p{0.6\textwidth}| } 
 \hline
\textbf{Ref}	&\textbf{Description}	                        &\textbf{Designed Goals}&	\textbf{Results (simulated)}\\ \hline
\cite{55_razaq2021privacy}          &Privacy-aware	&Security	& 
                            Their algorithm with security considerations performs better
                            than but close to the optimization values of ILP model; the number of offloaded
tasks with high-security levels is suppressed, while increased with medium- and low-security levels.
\\ \hline

\cite{42_guo2020blockchain}        & Stackelberg           & Security and QoS(Communication cost) & Commpare with PECRM, their design obtain 6.8\% more profit.   \\ \hline

\cite{44_an2022joint}              &Task Dependency	        &QoS(Communication and computation resources)	&  In both Slow-Fading and Fast-Fading Channels, their design of offloading outperforms the previous methods at total energy consumption, and their energy consumption also decreases. \\ \hline

\cite{56_jiang2022joint}          &Energy-Constrained      &QoS (Quality of experience)	& JORA-C performs better than JORA-D at each time slot, reaching the close-optimal solution. But, JORA-D needs less running time and achieves better performance in a high-density scenario. 

\\ \hline

\cite{57_chen2021dnnoff}            &DNNOff	        &QoS (Responsive time)	&Save Response time 12.4–66.6\%.\\ \hline
\cite{58_liu2023toward}                 &DNN-based Task	&QoS (acceleration ratio and service ratio)	& The average service ratio increases, and the average service ratio improved at 7\% to around 90\%\\ \hline
\cite{59_xu2020intelligent}         &Intelligent offloading	&Security 	& IOM keeps a high occupancy rate, lower energy consumption, decrease in propagation time, performs well in execution time.IOM is significantly better in completion time, and it maintains a high occupancy rate, and keeps higher privacy entropy. \\ \hline
\cite{60_dai2022task}               &Task co-offloading	    &QoS (Minimize system cost including task delay,  migration cost)	& When computation demands grow, it incurs less system cost; when data bits enlarges, it intends to avoid network congestion;it performs much better than the D2D algorithm; it shows superiority in system cost and learning regret.\\ \hline
\cite{61_hussein2020efficient}      &Efficient task	        &QoS (Response times and load of work balance)	& ACO algorithm reaches the best value in a reasonable time, makes the best offloading decision,  than the PSO algorithm. ACO task offloading algorithm enhanced the standard deviation than the RR algorithms. AOC can efficiently balance the workload over the Fog nodes.
\\ \hline
\cite{62_deng2021intelligent}       &Delay-aware partial	&QoS (Minimize delay)	&Reduces the delay by 23\%(Q-learning) and 30\%(DDPG). \\ \hline
\cite{63_wu2020accuracy}            &DNN inference	        &QoS (Accuracy, and latency)	& The accuracy  converges after Epoch = 1000; it outperforms benchmark schemes;it
guarantees error probability < 0.5\%; it can effectively reduce average service delay.
\\ \hline
\cite{64_tu2022task}                &LSTM prediction	    &QoS (Accuracy, response time)	&Reduce: latency(6.25\%),offloading cost(25.6\%), task throw rate(31.7\%).\\ \hline
\cite{65_galanopoulos2020improving} &IoT analytics	        &QoS (Computation resource)	& OnAlgo
consumes 60\% less power than ATO, while achieves 28\%~32\% accuracy than ATO and RCO for the case Bn = 0.005mW and H = 200MHz; OnAlgo exceeds both ATO and RCO by 5\% for the case Bn = 0.01mW and H = 2GHz; OnAlgo outperforms both ATO and RCO by 10\%~25\% for a large number of users.  \\ \hline
\cite{66_sun2021cloud}              &Cloud-edge	            &QoS (Computation resource)	&The accuracy stabilizes at 94.32\%;the task average delay is reduced by about 30\% than EPS;  the proposed scheme achieves the best results in task average delay and offloading failure rate.\\ \hline

\cite{68_liu2020cooperative}        &UAV-enabled	        &QoS (Network resource)	&  The centralized coop-UEC method achieves higher network utility, and lower drop rate,  than distributed coop-UEC method in terms of time-slot and the number of UAVs.\\ \hline

 \hline
\end{tabular}
\end{center}

\end{table*}

\subsubsection{Security} 

In EC, time-sensitive applications necessitate efficient accommodation of data by offloading computation tasks to nearby EC servers. While privacy concerns associated with data uploading to CC are neglected, the user privacy embedded in the vast data generated by the IoT and its associated computational tasks has not been adequately addressed, especially in the context of insecure offload procedures like data tampering, private data leakage, and data replication. For instance, utilizing an EC server with low-security credit by assigning data may be harassed by malicious use. The studies are categorized into: 1) Blockchain-based security and 2) Security Level.  
\begin{itemize}
\item Blockchain-based Security: Addressing data integrity and security becomes challenging as the volume of data escalates, particularly within the EC context where multiple trusted central entities coexist, and more offloading processes are needed. Initially, a blockchain is a decentralized public ledger that stores transaction records. Its principle is that each newly generated blockchain record is assigned a hash value by joining the previous block to the current block, and forms an irreversible chain. In the event of an attack on the management entity of the integrity-preserving system, the entire system is susceptible to threats. Blockchain is a candidate technology for such a decentralized security requirement. Reference \cite{42_guo2020blockchain}, they build a collaborative mining network by integrating EC-CC and non-mining devices to distribute mining tasks from the mobile blockchain to overcome the communication cost of mining processing. They use Stackelberg game to build the model of determining the optimal resource price and demands.

\item Security Level: Though Blockchain is frequently cited in network studies, this section covers the studies skipping the Blockchain technique. In \cite{55_razaq2021privacy}, offloading the smaller segments, divided from users’ IoT computation tasks to multiple Fog nodes is based on their security requirements, namely security level. The assigned Fog nodes perform the divided fragments collaboratively to eliminate the possibility of leaking data. Then, the fragments are flexibly offloaded to the CC server Fog server considering the requirements of deadline, security level, and resource usage. They also merge an integer linear programming model with a dynamic programming algorithm to maximize the number of IoT tasks under the constraints of security requirements and end-to-end transmission delay.
\end{itemize}

\subsubsection{Quality of service (QoS)}

In EC, rather than follow the QoS definition cited from the Internet context, we define the QoS with the goals of 1) reducing task latency, 2) achieving better resource utilization to parallel more tasks, and 3) increasing model accuracy.  The reduced latency is rooted in the intention of EC. Better resource utilization generally means more tasks in parallel. Obtaining a model with higher accuracy is the goal of the AI algorithms. 

\begin{itemize}
\item Lower Task Latency or Delay: Shorter latency is one of the preeminent research focuses on the topic of offloading because large latency-sensitive computing service tasks can be offloaded by multiple radio access technologies simultaneously. To achieve the goal, there are several considerations in researchers’ design. In \cite{61_hussein2020efficient}, the ant colony optimization (ACO) scheduler, an optimization meta-heuristic, is proposed to reduce the latency of IoT applications. Fog providers can maximize their benefits. Their design also takes into account the deadline constraints of IoT tasks. In \cite{57_chen2021dnnoff}, a novel approach to 1) extracting the structure of a DNN model, 2) converting the DNN model to a target program to process, offload, and deploy on target devices, 3) synthesizing an optimized offloading scheme involves the execution of various segments of the target program on suitable locations, namely DNNOff, is proposed to configure the offloading strategy based on on-demand offloading tasks. In \cite{62_deng2021intelligent}, to optimize the delay performance, they introduce two RL-based offloading strategies: 1) a Q-learning algorithm to determine a partial offloading decision, and 2) a deep deterministic policy gradient (DDPG). Their simulation indicates that the Q-learning model achieves a 23\% reduction in latency, while the DDPG scheme achieves a 30\% reduction in latency. In \cite{63_wu2020accuracy}, a deep RL-based algorithm is proposed to solve the Markov decision process for obtaining the optimal EC resource allocation solution, reducing the average service delay, and preserving long-term inference accuracy. In \cite{60_dai2022task}, a co-offloading framework is proposed in D2D-assisted MEC networks to minimize system costs, e.g., migration cost and task delay. Their framework can determine the optimal edge server without accessing the task’s offloading information. In \cite{56_jiang2022joint}, they proposed a framework called joint offloading and resource allocation, in centralized and distributed versions, to obtain a close-to-optimal performance within the constraint of 1) energy consumption and 2) computation and wireless resources.

\item Maximize task completions by better-leveraging resources: In the context of EC, layered architecture distributes different resources among EC architectures, making it challenging to leverage all the resources. Many studies focus on arranging tasks among EC to maximize resource utilization.  Partitioning in this scenario can be categorized as below. 

Leverage as many resources as possible: In \cite{65_galanopoulos2020improving}, they propose an algorithm first facilitating local execution on IoT devices. If its intelligent utilization anticipates a substantial enhancement in performance, it outsources to cloudlet servers. Reference \cite{61_hussein2020efficient} proposes two schedulers based on evolutionary algorithms, named ant colony optimization (ACO) and particle swarm optimization (PSO). The ACO scheduler attempts to maximize the profits of a Fog service constrained by the deadlines of the IoT tasks. The PSO scheduler is proposed to effectively load the balances among Fog nodes. Both cooperate to balance the communication cost and response time of IoT tasks among Fog nodes. In \cite{66_sun2021cloud}, a joint offloading scheme is proposed, utilizing resource occupancy prediction to formulate an optimal solution for task offloading in the presence of constrained edge resources. Different from computation resources, in \cite{68_liu2020cooperative} a DRL-based algorithm is introduced to maximize the long-term utility of the proposed UAV-enabled MEC network.

\item Better service accuracy: The goal of the DNN model is to offer as high-accuracy service as possible; thus, some research prefers accuracy as a new metric rather than latency. Under the EC environment, the limited and heterogeneous computing capability may shrink the service accuracy on edge servers, though many time-sensitive applications can tolerate lower accuracy by early exit \cite{70_dong2022multi}\cite{71_baccarelli2021learning}. In \cite{58_liu2023toward}, they propose a Distributed Partitioning and Offloading Solution, which designs an offloading strategy based on the task's priority and utilizes a stacking-based partitioning strategy,  to maximize
the acceleration ratio and service ratio. They use the DNN-based model and task partitioning in vehicular edge computing to maintain a reliable system. In \cite{64_tu2022task}, using Deep Reinforcement Learning (DRL) and Long Short-Term Memory (LSTM) networks, an Online Predictive Offloading (OPO) algorithm is proposed to determine the task offloading strategy, during the training process, improving the convergence accuracy along with the convergence speed.

\end{itemize}

\subsection{DNN Model Partitioning}

The constrained computational capacity of IoT devices impedes the local execution of DNN. As mentioned above, deep learning models naturally have a multi-layered structure and can be mapped into the components among IoT-EC-CC components. For instance, we can allocate the components with rich resources to training because the training phase needs more resources than the inference phase. Consequently, offloading selected layers of the DNN model to an EC server is a feasible strategy to alleviate the computational burden on IoT devices. The strategy of offloading is contingent upon the prevailing network conditions. Favorable network conditions permit the offloading of more DNN layers in shorter latency. After a reliable connection through a robust network is established, the DNN model inference can be partitioned and distributed among EC servers. 

In Table V, we list recent studies related to DNN models covering partitioning techniques, most of which are based on IoT applications. It also lists locations to execute tasks and corresponding resource constraints for using DNN partitioning. In summary, there are two aspects of partitioning relevant to DNN in IoT applications: 1) partial offloading methods of Deep Learning tasks; and 2) partitioning DNN models according to layers. In DNN partitioning, the partial offloading methods are similar to the task offloading mentioned for non-DL models. It first divides the application into a sequence of subtasks and then designs an offloading strategy to allocate resources to each subtask according to their requirements like energy and delay. On the other hand, since the DNN model has a layered structure, there is much research focusing on how to arrange the continuous subtasks distributed among the layers. We categorize the DNN model partitioning into 3 types: 1) DNN model partitioning for training, 2) DNN model partitioning for inference, and 3) Hybrid covering two different phases. In Table V, since DNN partitioning in training is much more difficult than the inference phase, research papers focusing on the training phase are much less available than ones on the inference phase. 

\subsubsection{DNN Model Partitioning for Training Phase}
Since DNN training needs much more computational resources than the inference phase, usually beyond the resources of the EC server, there are less papers focusing on DNN partitioning in the training phase. In \cite{5_ding2023resource}, the authors are driven by both data privacy and computational resource allocation on EC servers. They argue that in the training phase,  the way DNN model partitioning and privacy measurements are essential in defending against input reconstruction attacks. In \cite{7_yao2021context}, the authors introduce a framework across EC and CC servers by parallelizing the forward and backward steps of DNN training. They also parallelize computation and data transfer to avoid idles at EC servers and CC servers.

\begin{table*}[htbp]
\small
\centering
\caption{Works discussing DNN Model Partitioning}

\begin{center}

\begin{tabular}{ |p{0.03\textwidth}|p{0.2\textwidth}|c|p{0.3\textwidth}|l| } 
 \hline
\textbf{Ref}  &	\textbf{Description} &	\textbf{Running on}&	\textbf{Factors driving partitioning}	&\textbf{ \makecell[l]{Phases of DNN\\ model partitioning}} \\ \hline
\cite{5_ding2023resource} & DNN Partitioning & \makecell{Edge server} & \makecell[l]{Way of DNN partition, \\ computational resources} & DNN Training \\ \hline
\cite{7_yao2021context} & Context-Aware Compilation & \makecell{Edge server\\ Cloud server} & \makecell[l]{Data privacy at Edge, \\computational resources} & DNN Training \\ \hline
\cite{72_karjee2022split} &Split computing	&\makecell{Edge server\\IoT devices}	&\makecell[l]{Resource(computation  \\power and\\ communication, battery\\ load balance)}	&DNN Inference\\ \hline
\cite{73_kim2023partition} &Partition placement	&\makecell{Edge server\\IoT devices}	&\makecell[l]{Heterogeneous IoT Devices, \\computation power, battery} 	&DNN Inference\\ \hline
\cite{74_dong2021joint} &DNN Partitioning	&\makecell{Edge server  \\Cloud server}	& Model partitioning points	& DNN Inference\\ \hline
\cite{76_li2021throughput} &Delay-aware DNN	&Edge server	&Computation power	&DNN Inference\\ \hline
\cite{78_fang2023joint} &Joint Architecture	&IoT devices	&\makecell[l]{Computation resources, \\internet resources}	& DNN Inference\\ \hline
\cite{79_qu2023stochastic} &RL-Aided Adaptive	&\makecell{Edge server  \\IoT devices}	&\makecell[l]{Computation resource, \\communication resources}	&DNN Inference\\ \hline

\cite{81_yun2022cooperative} &Cooperative inference	&\makecell[l]{Edge Server \\ IoT devices}	&Resource(Memory)	&DNN Inference\\ \hline
\cite{82_samikwa2022adaptive} &Early exit 	&\makecell{Edge server  \\IoT devices}	&\makecell[l]{Resources(computation, \\communication, energy\\ consumption)}	&DNN Inference\\ \hline
\cite{83_su2022joint} &Resource allocation	&\makecell{Edge server  \\Cloud server}	&Resources(computation)	& Hybrid\\ \hline

\cite{80_zawish2022towards} &Resource-aware DNN 	&\makecell{Edge server  \\Cloud server}	&\makecell[l]{Resources(computation \\resource, flop, energy, etc.)}	& Hybrid\\ \hline
\cite{77_fan2023joint} &Joint DNN	&\makecell{Edge server\\Cloud server  \\IoT devices}	&\makecell[l]{Computation power, \\communication resource}	& Hybrid\\ \hline

 \hline
\end{tabular}

\end{center}

\end{table*}

\subsubsection{DNN Model Partitioning for Inference Phase}
In DL or traditional ML, the Inference task is to predict unseen datasets. It is the final but valuable step in AI generalization from seen data to unseen data after the model training task is completed. The inference task requires comparatively fewer computation resources than the training process and could be split into low-capacity devices. In \cite{72_karjee2022split}, an offloading strategy is proposed for the DNN model inference task. Their design owns an optimal partition plan covering an IoT device to an EC server. In \cite{73_kim2023partition}, DNN partition placement and resource allocation strategy are proposed to be determined by resources of different processing powers, memory, etc., among IoT or edge servers. In \cite{74_dong2021joint}, they propose a strategy for determining DNN model partitioning points for DNN inference. Their strategy uses heterogeneous resources to ensure the upper bound of inference delay. They also use Lyapunov optimization framework to reduce resource rental costs by up to 30\%.  In \cite{76_li2021throughput}, a novel model to address delay-aware DNN inference is introduced. Their objective is to maximize the amount of DNN service by expediting each DNN inference procedure. This is achieved through the cooperation of DNN partitioning and multi-thread parallelism. The cooperation spans both offline and online contexts of inference requests. In \cite{78_fang2023joint}, a framework advocating the execution of DNN inference in a partitioned and distributed structure across a sequence of IoT devices is introduced. This is achieved through model compression and the distribution of inference workloads based on adaptive workload partitioning, focusing on preserving accuracy. In the study \cite{79_qu2023stochastic}, an adaptive offloading decision approach employing reinforcement learning (RL) is applied. Their approach relies on a resource allocation solution for calculating rewards. This paper concentrates on classification tasks and explores methods to improve its delay performance and confidence level during DNN’s inference step through collaborative efforts from devices and the edge. Initially, they introduce a stochastic cumulative DNN inference scheme by merging multiple random DNN inference outcomes. The cumulative DNN formed is used to improve the confidence level. Then, employing a computationally efficient DNN model deployment strategy that involves shared computation between a locally deployed fast DNN model and a full DNN model partitioned between the IoT devices and EC server, they establish a closed-loop adaptive collaboration scheme for device-edge interaction to facilitate cumulative DNN inference across multiple devices. In \cite{81_yun2022cooperative}, they propose a memory-aware cooperative DNN inference to obtain high-performing lightweight DNNs with minimal delay. In \cite{82_samikwa2022adaptive}, an adaptive, energy-efficient, low-latency inference scheme is introduced to minimize both prediction latency and energy consumption.

\subsubsection{Hybrid of DNN Model Partitioning for Training Phase and Inference Phase}
In \cite{83_su2022joint}, their research minimizes the long-term average end-to-end delay, by collaborating DNN partitioning and computing resource allocation. Simultaneously, the objective is to guarantee that the energy consumption of both the EC and central CC does not exceed their designated energy budgets. Employing reinforcement learning and the Lyapunov optimization technique, a novel algorithm, referred to as Deep Deterministic Policy Gradient-based DNN Partition and Resource Allocation (DDPRA) is introduced. This algorithm is specifically designed to train a policy dynamically, utilizing observations from the environment to make informed decisions regarding DNN partitioning. In \cite{80_zawish2022towards}, collaborative DNN inference across EC and CC platforms is emerging as an effective strategy for facilitating various IoT applications. Edge devices, characterized by resource constraints, are unable to afford the workload inherent in DNN models. Consequently, researchers have adopted a distributed computing paradigm, involving the partitioning of a DNN model between the EC servers and the CC servers. In \cite{77_fan2023joint}, their focus lies in the context of the multi-base station to run EC server, and multi-service EC-CC. They collaborate to aid IoT devices in handling various DL tasks through task offloading.

\section{APPLICATIONS OF PARTITIONING IN IoT-EC-AI}

\noindent Before diving into the details of partitioning techniques in IoT-EC-CC-AI, several applications utilizing partitioning techniques are listed in this section. By now, many applications like AR/VR, and smart grid, driven by IoT-EC-CC-AI have been put into practice. In this section, we give a summary of these applications in Table VI. Most of them cover IoT-EC-CC and different AI algorithms. These applications are categorized into different application types, including AR/VR, Video analytics, Autonomous vehicles, and Smart Grid. Their Partitioning techniques used are also marked.

\begin{table*}[htbp]
\small
\centering
\caption{Applications working in IoT-EC-AI environment }
\begin{tabular}{p{0.1\textwidth} p{0.1\textwidth} p{0.02\textwidth} p{0.02\textwidth} p{0.02\textwidth} p{0.1\textwidth} p{0.1\textwidth} p{0.1\textwidth} p{0.1
\textwidth}  }
    \toprule
    \textbf{Ref} & \textbf{Description} &\textbf{IoT} &\textbf{EC} & \textbf{AI} & \textbf{Application}   & 
      \multicolumn{3}{c}{\textbf{Type of Partitioning Use}}\\
      \cmidrule(r){7-9}
     &  & & & &  & {Data}  & {Computation Task} & {DNN Model} \\
     &&&&&& {Partitioning} & {Partitioning} &{Partitioning} \\ \hline

    \cite{87_zhou2021deep} & Deep-learning-enhanced &\checkmark	&\checkmark	&\checkmark	&Video Analytics	    &\checkmark  &\checkmark	&	\\ \hline
    \cite{84_li2021cvc} &CVC &	&\checkmark	& \checkmark	&Video Analytics	    &\checkmark	&\checkmark	& \checkmark \\ \hline
    \cite{90_khayyam2020artificial} &Autonomous vehicles &\checkmark	&\checkmark	&\checkmark	&Autonomous Vehicles	&\checkmark	&\checkmark	& \\ \hline
    \cite{91_tang2020lopecs} &LoPECS &\checkmark	&\checkmark	&\checkmark	&Autonomous Vehicles	&\checkmark	&\checkmark	&\checkmark \\ \hline
    \cite{123_li2021adaptive} &Computing Scheduling &\checkmark	&\checkmark	&\checkmark	&Autonomous Vehicles	&\checkmark	&\checkmark	&\ \\ \hline
    \cite{93_abir2021iot} & Energy grid&\checkmark	&\checkmark	&\checkmark	&Smart Grid	            &\checkmark	&\checkmark	& \\ \hline
    \cite{94_hou2020p2p} &P2P network&\checkmark	&\checkmark	&	&Smart Grid	            &\checkmark	&\checkmark	& \\ \hline
    \cite{114_su2021secure} &Efficient Federated&\checkmark	&\checkmark	&\checkmark	&Smart Grid	            &\checkmark	&\checkmark	& \\ \hline
    
  \end{tabular}
\end{table*}

\subsection{Video Analytics}

The EC-CC paradigm presents a solution for the localized handling of the incessantly generated substantial volume of surveillance data within IoT systems \cite{87_zhou2021deep}. they proposed a system called A-YONet, a time-sensitive EC-CC surveillance system to detect multi-suspicious targets. Compared with the YOLOv3, faster R-CNN, and SSD, Y-YONet obtains better precision when detecting multiple dynamic objects in their experiment. Concurrently, time-sensitive video analytics plays a pivotal role in public safety applications, encompassing functions such as violence detection. It is also the technology foundation of self-driving, VR/AR, etc. Since surveillance cameras are booming, stream data uploading and downloading might exhaust the network bandwidth rapidly. In this scenario, the EC structure is beneficial for applications such as traffic management applications, crime detection, face recognition, etc. Deep learning can be utilized to produce high accuracy in these applications.  In their design, data are collected from cameras and trained at EC servers. Although deep learning is employed, they did not stress the DNN partitioning. 

In \cite{84_li2021cvc}, they propose an EC-based video caching framework to reduce latency and increase cache hit rate based on federated learning. They also design a collaborative cache decision method and a collaborative service response method to reduce hit cost.

\subsection{Autonomous Vehicles}

Many implicit or explicit environmental factors impact driving safety, including weather, illumination, road conditions, lane marking, etc.  In 2023, however, the National Highway Transportation Safety Administration (NHTSA) claims that the percentage of car accidents caused by human error ranges from 94\% to 96\%, including speeding, Distracted Driving, Improper Lookout, decision error, etc. To adapt to divergent driving conditions, autonomous vehicles have become a promising solution \cite{90_khayyam2020artificial}, with the advent of AI to minimize human intervention \cite{89_haydari2020deep}, sensors, and high-performance chips.

In Reference \cite{91_tang2020lopecs}, they propose a Low-Power Edge Computing System for real-time autonomous robots and vehicle services based on the cooperation of EC servers and CC servers. Their algorithm can determine the weight of task offloading dynamically constrained by the availability of CC nodes and optimal battery efficiency. In Reference \cite{123_li2021adaptive}, they propose an approach to determine the processing order among vehicles with the help of edge servers. Data are generated from each individual vehicle and sent to proximate edge nodes to process the request considering their different priorities.

\subsection{Smart Grid}

Generally, a traditional power grid system is composed of a mass of loosely interconnected synchronous Alternate Current (AC) grids, distributed in different electrical energy sections of generation, transmission, and distribution. Once the grid system is established, the electricity can be sent only in one direction, from a provider to the end-users. After the 1970s, conventional power grids underwent a metamorphosis into smart grids to rectify challenges within the prevailing power system, such as energy inefficiency, escalating energy demand, and concerns related to reliability and security. This transformation aimed to accommodate the diverse requirements of electronic devices, including electric vehicles \cite{86_ahmed2023dynamic}\cite{92_al2021adaptive}. Leveraging the capabilities of IoT devices, the traditional grid system can transform an efficient and intelligent energy grid, making use of the sensing, communication, and computing technologies inherent in IoT devices \cite{93_abir2021iot}. The challenge they face revolves around efficiently processing the substantial volume of data generated and uploaded from IoT devices. 

In \cite{94_hou2020p2p}, a P2P network-based EC smart grid model is proposed so that EC servers are connected in the P2P model. EC servers are responsible for collecting, computing, and storing data. Obviously, the data and tasks are distributed among EC servers. Their design promotes energy utilization and real-time control. In \cite{114_su2021secure}, they design a federate learning scheme for Artificial intelligence of things (AIoT) to preserve data privacy under a smart grid system. Their design can promote communication efficiency while keeping sharing local model updates and improve communication efficiency.

\section{FUTURE RESEARCH DIRECTIONS}

\noindent In our ongoing exploration of emerging network computing paradigms that harness the resources of IoT, EC Servers, and CC Servers, we recognize partitioning is applicable for scalable IoT applications for their different concerns. While extensive studies have investigated various network computing paradigms under different assumptions, this infrastructure, encompassing end devices, EC servers, and CC servers, is still in its infancy and requires further development to fuel the growing demand for IoT applications. In this section, we outline several promising directions for the partitioning techniques, based on the challenges for future research. We believe these directions are related to the partitioning technique and need more effort for future research. 

\begin{table*}[htbp]
\small
\centering
\caption{Future Research Directions for Partitioning }
\begin{tabular}{p{0.05\textwidth} p{0.3\textwidth}  p{0.15\textwidth}  p{0.15\textwidth} p{0.15\textwidth} }
    \toprule
    \textbf{Ref} & \textbf{Description}   &  \multicolumn{3}{c}{\textbf{Type of Partitioning Use}}\\ 
    \cmidrule(r){3-5}
     &  &  {Data}  & {Computation Task} & {DNN Model} \\ 
     &  & {Partitioning}  & {Partitioning}     & {Partitioning} \\ \hline
    \cite{101_tang2022double}\cite{106_samanta2021fault} &  Subtask &	&\checkmark  &   \\ \hline		
    \cite{102_tom2022functional} & Lightweight Virtualization & \checkmark  &\checkmark	&\\ \hline
    \cite{103_fu2022adaptive}  & APIs&\checkmark  &\checkmark	&	\\ \hline
    \cite{119_marzari2021towards}  & DRL& &\checkmark	&	\checkmark \\ \hline
    \cite{120_mishra2023transforming}  & Lightweight DL &  &	&\checkmark	\\ \hline

  \end{tabular}
  
\end{table*}

\subsection{Subtask Property for Computation Task Partitioning}

Computation task partitioning is a strategic tradeoff employed to balance workload distribution across different IoT-EC-CC components, a task determined by their varying resources and application requirements. Subtasks, a key concept, involve breaking down larger computational tasks into smaller, more manageable units. This approach facilitates the allocation and offloading of these subtasks to edge devices or servers, leading to a more efficient and distributed processing method. The computation task partition technique optimizes system performance, resource utilization, latency reduction, and scalability. The subtask property plays a crucial role in determining the computation task partitioning strategy and enables effective collaboration among edge devices, ensuring optimal resource utilization. Investigating subtask requirements and available architecture resources is crucial in this context. For instance, reference \cite{101_tang2022double} proposes a Dynamic Framing Offloading algorithm that divides vehicle tasks into refined sequential subtasks, achieving superior offloading performance by minimizing total delay and waiting time. Additionally, successful offloading relies on robust network communication. A common issue is the failure to establish communication and connectivity among IoT devices. One solution that has been proposed is to design a tolerant offloading model \cite{106_samanta2021fault}.

\subsection{Lightweight and Distributed AI Models for Time-sensitive IoT Services}
Compared with CC servers,  IoT devices, and EC servers have limited resources, making it challenging to support large machine learning models and data analytics algorithms. Often, the workload needs to be deployed on, trained on, and downloaded from CC servers. To leverage resources among mobile and IoT devices, there's a need to explore lightweight and distributed ML algorithms that require less energy consumption, storage, and computation requirements and can be distributed across the network. This is crucial not only for resource allocation but also for time-sensitive applications. Ref \cite{120_mishra2023transforming} also reviews five different types of methods to compress DNN including network pruning, sparse representation, bits precision, knowledge distillation, and miscellaneous. 

\subsection{Lightweight Virtualization for Heterogeneous IoT Structures}

IoT hardware's lightweight and heterogeneous nature presents unique challenges for IoT applications. These challenges include integrating different types of data and customized operating systems, thereby decreasing the flexibility of service sharing. In response to these challenges, lightweight virtualization techniques, such as containers, have gained prominence. Containers effectively monitor resource consumption within containers and provide notifications for necessary interventions. Containers themselves work for data partitioning. For example, reference \cite{102_tom2022functional} claims the promising candidates for lightweight virtualization are containers and unikernels. Light virtualization aligns well with the limited resources of IoT devices, which qualifies it as a crucial technology for building scalable and manageable IoT ecosystems.

\subsection{Standardized APIs}

Microservices and APIs can be used in EC for IoT applications. The Microservices architecture allows complex IoT applications to be decomposed into smaller, independently deployable services. These microservices can expose APIs, enabling communication among various IoT devices and edge components. APIs serve as bridges for data sharing, action triggering, and task coordination within the edge network. This approach enhances the flexibility, scalability, and agility of IoT systems, making them adaptable to changing requirements. As the IoT ecosystem evolves, microservices and APIs in EC become essential tools for building responsive, distributed, and interconnected IoT applications. The microservice mapper partitions the microservice graph of services into multiple parts and maps them
to the edge nodes and cloud nodes. For instance, microservices can be deployed into correct nodes of the cloud-edge continuum \cite{103_fu2022adaptive}. Thus, more research efforts are needed on API standardization.

\subsection{Deep Reinforcement Learning(DRL) Algorithm in Designing Task Partitioning Policy }

Deep Reinforcement Learning plays a pivotal role in EC, particularly in determining the computation task partitioning strategy. By integrating DNN models at the edge servers, it becomes possible to analyze data locally and make informed decisions regarding the computation task partitioning. For instance, DRL policy trains subtasks independently with a sub-goal-directed reward function for each subtask\cite{119_marzari2021towards}. It takes subtasks as parameters in their proposed algorithm. This capability empowers edge devices to assess the computational requirements of various subtasks. This assessment, coupled with time-sensitive data processing capabilities, allows edge devices to dynamically select subtasks for computation task partitioning, optimizing resource utilization, and minimizing latency. Deep learning models continuously adapt to changing conditions, ensuring that the computation task partitioning strategy remains efficient and responsive to the evolving demands of EC applications.

\section{CONCLUDING}

First, with the surge of IoT market opportunities, IoT applications propose new requirements of real-time data transition and powerful capability of data processing to traditional cloud network architecture. Though CC offers high-performance computation capability as a desired tool for processing data, however, transmission latencies inherent in the traditional cloud architecture is not supportive of time-sensitive applications. EC is a promising computing technique for IoT with the help of the partitioning technique. EC is introduced for time-sensitive applications by reducing data transmission through the Internet. However, the implementation of EC with AI for IoT applications poses significant challenges. Secondly, we dived more into the architecture of IoT, EC, and DNN models. From the perspective of layered structures of EC, and DNN models, the partitioning technique becomes a valid and essential technique for IoT-EC-AI. From the perspective of data security, partitioning techniques can also reduce data privacy concerns. Thirdly, we have investigated various partitioning technologies in three directions: data partitioning, computation task partitioning, and DNN model partitioning. We investigate data partitioning for security reasons, computation task partitioning for security and QoS purposes, and DNN model partitioning in different phases. Then, we investigate different applications from the partitioning perspective by connecting them to related partitioning techniques. Finally, we point out five related and inadequate research directions focusing on the partitioning technique. We believe these directions can promote the partitioning technique to a further level for security and efficiency purposes.        

The emerging combination of IoT-EC-AI is changing our modern life rapidly and widely to promote the quality of life as well as the industry. Partitioning helps overcome challenges in IoT-EC-AI concerning data privacy, and efficient resource allocation. This survey gives insights into their concepts, architecture, requirements, advantages, and disadvantages. We hope this survey can elicit more focus on IoT-EC-AI research.

\section{APPENDIX}
\begin{table*}[htbp]
\centering
\caption{ACRONYMS/ABBREVIATIONS }

\begin{tabular}{  l | l  } 
 \hline
    \textbf{Acronym} & \textbf{Definition} \\ \hline
	\textbf{EC} & Edge Computing  \\ 
	\textbf{IoT} & Internet of Things \\ 
	\textbf{MEC} & Mobile-access Edge Computing \\ 
	\textbf{UAV} & Unmanned Aerial Vehicle \\ 
	\textbf{AR} & Augmented Reality\\ 
	\textbf{ML} & Machine Learning\\ 
	\textbf{DL} & Deep Learning \\ 
	\textbf{SBCS} & Smart Building Control System \\ 
	\textbf{VM} & Virtual Machine\\ 
	\textbf{CC} & Cloud Computing\\ 
	\textbf{MQTT} & Message Queuing Telemetry Transport\\ 
	\textbf{CoAP} & Constrained Application Protocol\\ 
	\textbf{HTTP} & Hypertext Transfer Protocol \\ 
	\textbf{LwM2M} & OMA Lightweight M2M \\ 
	\textbf{LoRaWAN} & Low Power Wide Area Networking (LPWAN) communication protocol that functions on LoRa \\
    
    \textbf{DNN} & Deep Neural Networks \\
 \hline
\end{tabular}
\end{table*}

\begin{wrapfigure}{l}{25mm} 
    \includegraphics[width=1in,height=1.25in,clip,keepaspectratio]{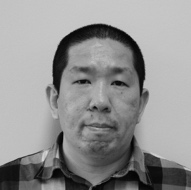}
  \end{wrapfigure}\par
  \textbf GUOXING YAO received the M.S. degree in Computer Science from Syracuse University in 2020. He is currently pursuing a Ph.D. in computer science at the University of Missouri–St. Louis, Missouri. His research interests include data mining, big data analytics, machine learning, and security.\par
  
\section*{}

\begin{wrapfigure}{l}{25mm} 
    \includegraphics[width=1in,height=1.25in,clip,keepaspectratio]{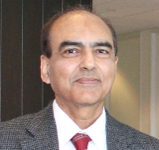}
  \end{wrapfigure}\par
  \textbf LAV GUPTA (M’86–SM’05) received a master’s degree in electrical engineering from IIT Kanpur, India in 1980 and MS and PhD in computer science and engineering from Washington University in St Louis, MO, USA in 2019. He is currently an Assistant Professor of computer science at the University of Missouri in St. Louis.
He has worked for about fifteen years in telecommunications planning, deployment, and regulation. He has also worked as a senior faculty of computer science and network planning in India and the UAE for several years.  He is the author of one book, 19 papers, and has been a speaker at many international conferences and seminars. His current research is at the intersection of cybersecurity, IoT-edge-cloud, AI and security of critical infrastructure and next generation cellular wireless networks. 
Dr. Gupta is a senior member of IEEE and a member of ACM. He has been a recipient of the best author award from Elsevier and has figured in the Stanford list of top 2\% researchers globally.
\par

\end{document}